\documentclass[twocolumn,aps,prc,superscriptaddress,showpacs,floatfix]{revtex4}
\usepackage{amsmath}
\usepackage{graphicx}

\begin{document}

\title{Light cluster production in intermediate energy heavy-ion collisions
induced by neutron-rich nuclei}
\author{Lie-Wen Chen}
\thanks{On leave from Department of Physics, Shanghai Jiao Tong University,
Shanghai 200030, China}
\affiliation{Cyclotron Institute and Physics Department, Texas A\&M University, College
Station, Texas 77843-3366}
\author{C. M. Ko}
\affiliation{Cyclotron Institute and Physics Department, Texas A\&M University, College
Station, Texas 77843-3366}
\author{Bao-An Li}
\affiliation{Department of Chemistry and Physics, P.O. Box 419, Arkansas State
University, State University, Arkansas 72467-0419}
\date{\today}

\begin{abstract}
The coalescence model based on nucleon distribution functions from an
isospin-dependent transport model is used to study the production of
deuteron, triton, and $^{3}$He from heavy-ion collisions induced by
neutron-rich nuclei at intermediate energies. It is found that the emission
time of these light clusters depends on their masses. For clusters with same
momentum per nucleon, heavier ones are emitted earlier. Both the yield and
energy spectrum of light clusters are sensitive to the density dependence of
nuclear symmetry energy, with more light clusters produced in the case of a
stiff symmetry energy. On the other hand, effects due to the stiffness of
the isoscalar part of nuclear equation of state and the medium dependence of
nucleon-nucleon cross sections on light cluster production are unimportant.
We have also studied the correlation functions of clusters, and they are
affected by the density dependence of nuclear symmetry energy as well, with
the stiff symmetry energy giving a stronger anti-correlation of light
clusters, particularly for those with large kinetic energies. Dependence of
light cluster production on the centrality and incident energy of heavy ion
collisions as well as the mass of the reaction system is also investigated.
\end{abstract}

\maketitle

\section{Introduction}

The equation of state (\textrm{EOS}) of an asymmetric nuclear matter with
unequal numbers of protons and neutrons depends crucially on the nuclear
symmetry energy. Although the nuclear symmetry energy at normal nuclear
matter density is known to be around $30$ \textrm{MeV} from the empirical
liquid-drop mass formula \cite{myers,pomorski}, its values at other
densities are poorly known. Studies based on various theoretical models also
give widely different predictions \cite{ibook}. Lack of this knowledge has
hampered our understanding of both the structure of radioactive nuclei \cite%
{oya,brown,hor01,furn02} and many important issues in nuclear astrophysics %
\cite{bethe,bom,lat01}, such as the nucleosynthesis during the pre-supernova
evolution of massive stars and the properties of neutron stars \cite%
{bethe,lat01}. However, recent advances in radioactive nuclear beam
facilities provides a unique opportunity to study the density dependence of
nuclear symmetry energy \cite{npa01,ibook,li98,ditoro99}. Theoretical
studies have already shown that in heavy-ion collisions induced by
neutron-rich nuclei, the effect of nuclear symmetry energy can be studied
via the pre-equilibrium neutron/proton ratio \cite{li97}, the isospin
fractionation \cite{fra1,fra2,xu00,tan01,bar02}, the isoscaling in
multifragmentation \cite{betty}, the proton differential elliptic flow \cite%
{lis}, the neutron-proton differential transverse flow \cite{li00,Greco03},
and the $\pi ^{-}$ to $\pi ^{+}$ ratio \cite{li02}. Also, it was found in a
recent work that the strength of the correlation function for nucleon pairs
with high total momenta in heavy-ion collisions induced by neutron-rich
nuclei is sensitive to the nuclear symmetry energy \cite{CorShort}. This
result demonstrates that the space-time structure of neutron and proton
emission functions is affected by the density dependence of nuclear symmetry
energy. Since deuteron production in heavy-ion collisions depends on the
space-time and momentum distributions of neutrons and protons at freeze out %
\cite{mrow92}, its production as well as that of other light clusters such
as triton and $^{3}$He are also expected to be sensitive to the density
dependence of nuclear symmetry energy.

To study this quantitatively, we have used in Ref. \cite{ClstShort} a
coalescence model for treating cluster production from an isospin-dependent
Boltzmann-Uehling-Uhlenbeck (IBUU) transport model. We found that both the
multiplicities and energy spectra of light clusters are indeed sensitive to
the density dependence of nuclear symmetry energy, with the stiff symmetry
energy giving a larger yield of clusters. In particular, the t$/^{3}$He
ratio at high kinetic energies show the strongest dependence on the
stiffness of nuclear symmetry energy. It was also pointed out in Ref. \cite%
{ClstShort} that the effects due to the isospin-independent part of nuclear
equation of state and the medium dependence of nucleon-nucleon cross
sections are unimportant for light cluster production in heavy ion
collisions induced by neutron-rich nuclei. In the present paper, we describe
in detail the model used for studying light cluster production in heavy-ion
collisions and present also additional results. In particular, we show the
correlation functions of these light clusters and the dependence of light
cluster production on the centrality and incident energy of heavy ion
collisions as well as the mass of the reaction system.

The paper is organized as follows. In Section \ref{coalescence}, we describe
the coalescence model for light cluster production from the nucleon
phase-space distributions at freeze out. This includes the construction of
the Wigner phase-space density functions for deuteron, triton, and $^3$He.
The isospin-dependent Boltzmann-Uehling-Uhlenbeck (IBUU) transport model
used for modeling heavy ion collisions is briefly described in Section \ref%
{ibuu}. Using the coalescence model given in the previous section, the IBUU
model is first applied to study light cluster production in heavy ion
collisions involving symmetric nuclei, and the results are compared with
available experimental data. Results on light cluster production from
heavy-ion collisions induced by neutron-rich nuclei at intermediate energies
are presented in Section \ref{results}. We show not only the multiplicities
and energy spectra of light clusters but also their dependence on the
stiffness of nuclear symmetry energy. Dependence of light cluster production
on the impact parameter, incident energy, and masses of the reaction system
is also given. In Section \ref{correlation}, we study correlations of light
clusters and their dependence on nuclear symmetry energy. We then conclude
with a summary and outlook in Section \ref{summary}.

\section{Cluster production from nucleon coalescence}

\label{coalescence}

Light cluster production has been extensively studied in experiments
involving heavy ion collisions at all energies (e.g., see Ref. \cite%
{Hodgson03} for a recent review). A popular model for describing the
production of light clusters in these collisions is the coalescence model
(e.g., see Ref. \cite{Csernai86} for a theoretical review), which has been
used at both intermediate \cite{Gyu83,Aich87,Koch90} and high energies \cite%
{Mattie95,Nagle96}. In most applications of the coalescence model, the
energy spectra for clusters are simply given by the product of the spectra
of their constituent nucleons multiplied by an empirical coalescence factor.
In Ref. \cite{Indra00}, for example, this simplified coalescence model based
on the nucleon distributions obtained from the intra-nuclear cascade model
gives a satisfactory description of measured high energy spectra of light
clusters in intermediate energy (about $100$ \textrm{MeV/nucleon}) heavy ion
collisions. In more sophisticated coalescence model, the coalescence factor
is computed from the overlap of the cluster Wigner phase-space density with
the nucleon phase-space distributions at freeze out. The coalescence factor
obtained from light particle emission spectra thus provides important
information about the space-time structure of nucleon emission source and is
thus a sensitive probe to the dynamics of heavy ion collisions \cite%
{Pearson63,Schwarz63,Gutbrod76,Sato81,Mekjian78,Heinz99}. In the present
study, this more microscopic coalescence model is adopted for studying light
cluster production from heavy ion collisions induced by neutron-rich nuclei
at intermediate energies.\vspace{-0.5cm}

\subsection{The coalescence model}

In the coalescence model, the probability for producing a cluster of
nucleons is determined by the overlap of its Wigner phase-space density with
the nucleon phase-space distributions at freeze out. For a system containing 
$A$ nucleons, the momentum distribution of a $M$-nucleon cluster with $Z$
protons can be expressed as \cite{Mattie95}%
\begin{eqnarray}
&&\frac{dN_{M}}{d^{3}K}=G\left( 
\begin{array}{c}
A \\ 
M%
\end{array}%
\right) \left( 
\begin{array}{c}
M \\ 
Z%
\end{array}%
\right) \frac{1}{A^{M}}\int \left[ \prod_{i=1}^{Z}f_{p}(\mathbf{r}_{i},%
\mathbf{k}_{i})\right]   \notag \\
&&\times \left[ \prod_{i=Z+1}^{M}f_{n}(\mathbf{r}_{i},\mathbf{k}_{i})\right]
\rho ^{W}(\mathbf{r}_{i_{1}},\mathbf{k}_{i_{1}}\cdots \mathbf{r}_{i_{M-1}},%
\mathbf{k}_{i_{M-1}})  \notag \\
&&\times \delta (\mathbf{K}-(\mathbf{k}_{1}+\cdots +\mathbf{k}_{M}))d\mathbf{%
r}_{1}d\mathbf{k}_{1}\cdots d\mathbf{r}_{M}d\mathbf{k}_{M}.  \label{Coal}
\end{eqnarray}%
where $f_{n}$ and $f_{p}$ are, respectively, the neutron and proton
phase-space distribution functions at freeze-out; $\rho ^{W}$ is the Wigner
phase-space density of the $M$-nucleon cluster; $\mathbf{r}_{i_{1}},\cdots ,%
\mathbf{r}_{i_{M-1}}$ and $\mathbf{k}_{i_{1}},\cdots ,\mathbf{k}_{i_{M-1}}$
are, respectively, the $M-1$ relative coordinates and momenta taken at equal
time in the $M$-nucleon rest frame; and $G$ is the spin-isospin statistical
factor for the cluster. In transport model simulations for heavy ion
collisions, the multiplicity of a $M$-nucleon cluster is then given by \cite%
{Mattie95} 
\begin{eqnarray}
N_{M} &=&G\int \underset{i_{1}>i_{2}>...>i_{M}}{\sum }d\mathbf{r}_{i_{1}}d%
\mathbf{k}_{i_{1}}\cdots d\mathbf{r}_{i_{M-1}}d\mathbf{k}_{i_{M-1}}  \notag
\\
&\times &\langle \rho _{i}^{W}(\mathbf{r}_{i_{1}},\mathbf{k}_{i_{1}}\cdots 
\mathbf{r}_{i_{M-1}},\mathbf{k}_{i_{M-1}})\rangle .  \label{MultCluster}
\end{eqnarray}%
In the above, $\langle \cdots \rangle $ denotes event averaging and the sum
runs over all possible combinations of $M$ nucleons containing $Z$ protons.
The spin-isospin statistical factor $G$ is $3/8$ for deuteron and $1/3$ for
triton or $^{3}$He, with the latter including the possibility of coalescence
of a deuteron with another nucleon to form a triton or $^{3}$He \cite%
{Polleri99}.

As pointed out in Ref. \cite{Mattie95}, the validity of coalescence model is
based on the assumption that nucleon emissions are statistically independent
and binding energies of formed clusters as well as the quantum dynamics only
play minor roles. Since the binding energies of deuteron, triton and $^{3}$%
He are $2.2$ \textrm{MeV}, $7.72$ \textrm{MeV} and $8.48$ \textrm{MeV},
respectively, the coalescence model for the production of these light
clusters in heavy-ion collisions is thus applicable if the colliding system
or the emission source has excitation energy per nucleon or temperature
above $\sim 9$ MeV. Furthermore, the coalescence model is a perturbative
approach and is valid only if the number of clusters formed in the
collisions is small. As shown later, this condition is indeed satisfied for
energetic deuterons, tritons, and $^3$He in the collisions we are
considering.

In the following, we construct the Wigner phase-space densities for
deuteron, triton, and $^3$He. The method can be generalized, although more
involved, to the Wigner phase-space density for four-nucleon clusters (e.g., 
$^{4}$He) and $M$-nucleon clusters.

\subsection{Wigner phase-space density of deuteron}

For deuterons, we obtain their Wigner phase-space density from the
Hulth\'{e}n wave function \cite{Hodgson71}, i.e., 
\begin{equation}
\phi (r)=\sqrt{\frac{\alpha \beta (\alpha +\beta )}{2\pi (\alpha -\beta )^{2}%
}}\frac{\exp (-\alpha r)-\exp (-\beta r)}{r}  \label{HulthenR}
\end{equation}%
with the parameters $\alpha =0.23$ \textrm{fm}$^{-1}$ and $\beta =1.61$ 
\textrm{fm}$^{-1}$ to reproduce the measured deuteron root-mean-square
radius of $1.96$ \textrm{fm} \cite{Ericsson88}. The Wigner phase-space
density of deuteron can then be obtained from 
\begin{equation}
\rho _{d}^{W}(\mathbf{r},\mathbf{k})=\int \phi (\mathbf{r+}\frac{\mathbf{R}}{%
2})\phi ^{\ast }(\mathbf{r-}\frac{\mathbf{R}}{2})\exp (-i\mathbf{k}\cdot 
\mathbf{R})d\mathbf{R},  \label{Wigner}
\end{equation}%
where $\mathbf{k=(k}_{1}\mathbf{-k}_{2}\mathbf{)}/2$ is the relative
momentum while $\mathbf{r=(r}_{1}\mathbf{-r}_{2}\mathbf{)}$ is the relative
coordinate of proton and neutron in the deuteron. The $\rho _{d}^{W}(\mathbf{%
r},\mathbf{k})$ cannot be derived analytically if one uses Eq. (\ref%
{HulthenR}). As in Ref. \cite{Mattie95}, we express the Hulth\'{e}n wave
function in terms of the sum of $15$ Gaussian wave functions 
\begin{equation}
\phi (r)=\sum_{i=1}^{15}c_{i}\left( \frac{2w_{i}}{\pi }\right) ^{3/4}\exp
(-w_{i}r^{2}),  \label{HulthenGau}
\end{equation}%
where the coefficient $c_{i}$ and the width parameter $w_{i}$ in the
Gaussian wave function are determined by least square fit, and they are
given in Table \ref{deuteron}. With the above deuteron wave function, $\rho
_{d}^{W}(\mathbf{r},\mathbf{k})$ can be obtained analytically as \cite%
{Shin90,Mattie95}%
\begin{eqnarray}
&&\rho _{d}^{W}(\mathbf{r},\mathbf{k})=8\sum_{i=1}^{15}c_{i}^{2}\exp \left(
-2w_{i}r^{2}-\frac{k^{2}}{2w_{i}}\right)   \notag \\
&&+16\sum_{i>j}^{15}c_{i}c_{j}\left( \frac{4w_{i}w_{j}}{(w_{i}+w_{j})^{2}}%
\right) ^{3/4}\exp \left( -\frac{4w_{i}w_{j}}{w_{i}+w_{j}}r^{2}\right)  
\notag \\
&&\times \exp \left( -\frac{k^{2}}{w_{i}+w_{j}}\right) \times \cos \left( 2%
\frac{w_{i}-w_{j}}{w_{i}+w_{j}}\mathbf{r}\cdot \mathbf{k}\right) .
\label{WignerD}
\end{eqnarray}

\begin{table}[tbp]
\caption{{\protect\small {Gaussian fit coefficient }$c_{i}$\ and $w_{i}$ to
deuteron wave function.}}
\label{deuteron}%
\begin{tabular}{p{0.5in}p{0.8in}p{0.8in}}
\hline\hline
$i$ & \quad $c_{i}$\quad & $\quad w_{i}(1/$\textrm{fm}$^{\mathrm{2}})$ \\ 
\hline
1 & 3.49665E$-$1 & 1.57957E$-$2 \\ 
2 & 1.85419E$-$1 & 3.94293E$-$2 \\ 
3 & 1.72279E$-$1 & 8.99793E$-$2 \\ 
4 & 4.62152E$-$2 & 9.75943E$-$2 \\ 
5 & 1.49458E$-$1 & 1.80117E$-$1 \\ 
6 & 7.74205E$-$2 & 1.93353E$-$1 \\ 
7 & 1.48268E$-$4 & 1.99811E$-$1 \\ 
8 & 7.35549E$-$3 & 2.17921E$-$1 \\ 
9 & 4.89047E$-$2 & 2.89902E$-$1 \\ 
10 & 4.19816E$-$2 & 4.70739E$-$1 \\ 
11 & 1.72670E$-$2 & 4.89604E$-$1 \\ 
12 & 1.06294E$-$1 & 9.27621E$-$1 \\ 
13 & 2.51462E$-$4 & 1.98822E$+$0 \\ 
14 & 3.22947E$-$2 & 2.59243E$+$0 \\ 
15 & 1.15826E$-$2 & 1.44639E$+$1 \\ \hline\hline
\end{tabular}%
\end{table}

\vspace{0.2cm}

The expansion in Eq. (\ref{HulthenGau}) approximates the Hulth\'{e}n wave
function not only in coordinate space but also in momentum space. In
momentum space, the normalized Hulth\'{e}n wave function is given by 
\begin{equation}
\phi (k)=\frac{\sqrt{(\alpha +\beta )^{3}\alpha \beta }}{\pi (\alpha
^{2}+k^{2})(\beta +k^{2})}.  \label{HulthenP}
\end{equation}

\begin{figure}[th]
\includegraphics[scale=0.8]{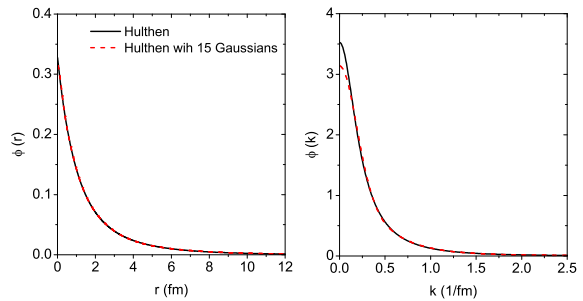}
\caption{{\protect\small The exact (solid curves) and fitted (dashed curves)
Hulth\'{e}n wave function in coordinate space (left panel) and momentum
space (right panel).}}
\label{HulthenFig}
\end{figure}

In Fig. \ref{HulthenFig}, we show the exact (solid lines) and fitted (dashed
lines) Hulth\'{e}n wave function in both the coordinate and the momentum
space. It is seen that the sum of $15$ Gaussian wave functions reproduces
satisfactorily the exact Hulth\'{e}n wave function both in coordinate and
momentum spaces.

\subsection{Wigner phase-space densities of triton and $^{3}$He}

For triton and $^{3}$He, their Wigner phase-space densities are obtained
from their internal wave functions, which are taken to be those of a
spherical harmonic oscillator \cite{Dover83,Heinz99}, i.e., 
\begin{equation}
\psi (\mathbf{r}_{1},\mathbf{r}_{2},\mathbf{r}_{3})=(3\pi
^{2}b^{4})^{-3/4}\exp \left( -\frac{\mathbf{\rho }^{\mathbf{2}}+\mathbf{%
\lambda }^{\mathbf{2}}}{2b^{2}}\right) .  \label{triWF}
\end{equation}%
In the above, we have used the normal Jacobi coordinates for a
three-particle system, i.e.,%
\begin{equation}
\left( 
\begin{array}{c}
\mathbf{R} \\ 
\mathbf{\rho } \\ 
\mathbf{\lambda }%
\end{array}%
\right) =J\left( 
\begin{array}{c}
\mathbf{r}_{1} \\ 
\mathbf{r}_{2} \\ 
\mathbf{r}_{3}%
\end{array}%
\right) ,  \label{Jacobi1}
\end{equation}%
where $\mathbf{R}$ is the center-of-mass coordinate, while $\mathbf{\rho }$
and $\mathbf{\lambda }\ $are relative coordinates. The Jacobian matrix is
given by 
\begin{equation}
J=\left( 
\begin{array}{ccc}
\frac{1}{3} & \frac{1}{3} & \frac{1}{3} \\ 
\frac{1}{\sqrt{2}} & -\frac{1}{\sqrt{2}} & 0 \\ 
\frac{1}{\sqrt{6}} & \frac{1}{\sqrt{6}} & -\frac{2}{\sqrt{6}}%
\end{array}%
\right) .  \label{Jacobi2}
\end{equation}%
>From Eqs. (\ref{Jacobi1}) and (\ref{Jacobi2}), one obtains $d\mathbf{r}_{1}d%
\mathbf{r}_{2}d\mathbf{r}_{3}=1/|J|^{3}d\mathbf{R}d\mathbf{\rho }d\mathbf{%
\lambda =}3^{3/2}d\mathbf{R}d\mathbf{\rho }d\mathbf{\lambda }$. It is then
easy to check that the wave function Eq. (\ref{triWF}) is normalized to one.
>From the relation $(\mathbf{r}_{1}-\mathbf{R)}^{\mathbf{2}}+(\mathbf{r}_{2}-%
\mathbf{R)}^{\mathbf{2}}+(\mathbf{r}_{3}-\mathbf{R)}^{\mathbf{2}}=\mathbf{%
\rho }^{\mathbf{2}}+\mathbf{\lambda }^{\mathbf{2}}$, the root-mean-square
radius $r_{\mathrm{rms}}$ of triton or $^{3}$He can be written as 
\begin{equation}
r_{\mathrm{rms}}^{2}=\int \frac{\mathbf{\rho }^{\mathbf{2}}+\mathbf{\lambda }%
^{\mathbf{2}}}{3}|\psi (\mathbf{r}_{1},\mathbf{r}_{2},\mathbf{r}%
_{3})|^{2}3^{3/2}d\mathbf{\rho }d\mathbf{\lambda }=b^{2}.  \label{rms}
\end{equation}%
The parameter $b$ is determined to be $1.61$ fm and $1.74$ fm for triton and 
$^{3}$He, respectively, from their measured root-mean-square radii \cite%
{chen86}.

The Wigner phase-space densities for triton and $^{3}$He is then given by 
\begin{eqnarray}
&&\rho _{\text{t}(^{3}\text{He})}^{W}(\mathbf{\rho },\mathbf{\lambda },%
\mathbf{k}_{\mathbf{\rho }},\mathbf{k}_{\mathbf{\lambda }})  \notag \\
&=&\int \psi\left(\mathbf{\rho +}\frac{\mathbf{R}_{\mathbf{1}}}{2},\mathbf{%
\lambda +}\frac{\mathbf{R}_{\mathbf{2}}}{2}\right)\psi ^{\ast } \left(%
\mathbf{\rho -}\frac{\mathbf{R}_{\mathbf{1}}}{2},\mathbf{\lambda -} \frac{%
\mathbf{R}_{\mathbf{2}}}{2}\right)  \notag \\
&&\times \exp (-i\mathbf{k}_{\mathbf{\rho }}\cdot \mathbf{R}_{\mathbf{1}%
})\exp (-i\mathbf{k}_{\mathbf{\lambda }}\cdot \mathbf{R}_{\mathbf{2}%
})3^{3/2}d\mathbf{R}_{\mathbf{1}}d\mathbf{R}_{\mathbf{2}}  \notag \\
&=&8^{2}\exp\left(-\frac{\mathbf{\rho }^{\mathbf{2}} +\mathbf{\lambda }^{%
\mathbf{2}}}{b^{2}}\right) \exp (-(\mathbf{k}_{\mathbf{\rho }}^{2}+\mathbf{k}%
_{\mathbf{\lambda }}^{2})b^{2}),  \label{WignerTri}
\end{eqnarray}%
where $\mathbf{k}_{\mathbf{\rho }}$ and $\mathbf{k}_{\mathbf{\lambda }}$ are
relative momenta, which together with the total momentum $\mathbf{K}$ are
defined by%
\begin{equation}
\left( 
\begin{array}{c}
\mathbf{K} \\ 
\mathbf{k}_{\mathbf{\rho }} \\ 
\mathbf{k}_{\mathbf{\lambda }}%
\end{array}%
\right) =J^{-,+}\left( 
\begin{array}{c}
\mathbf{k}_{1} \\ 
\mathbf{k}_{2} \\ 
\mathbf{k}_{3}%
\end{array}%
\right) ,  \label{Jacobi4}
\end{equation}%
with $\mathbf{k}_{1}$, $\mathbf{k}_{2}$, and $\mathbf{k}_{3}$ being momenta
of the three nucleons. The matrix $J^{-,+}$ denotes the inverse of the
conjugate complex transposition of the Jacobian matrix $J$, i.e.,%
\begin{equation}
J^{-,+}=\left( 
\begin{array}{ccc}
1 & 1 & 1 \\ 
\frac{1}{\sqrt{2}} & -\frac{1}{\sqrt{2}} & 0 \\ 
\frac{1}{\sqrt{6}} & \frac{1}{\sqrt{6}} & -\frac{2}{\sqrt{6}}%
\end{array}%
\right) .  \label{Jacobi5}
\end{equation}

\section{IBUU model and light cluster production}

\label{ibuu}

To determine the space-time and momentum distributions of nucleons at freeze
out in intermediate-energy heavy ion collisions, we use the
isospin-dependent Boltzmann-Uehling-Uhlenbeck (IBUU) transport model \cite%
{li97,li00,li02,li96}. For a review of the \textrm{IBUU} model, we refer the
reader to Ref. \cite{li98}. The \textrm{IBUU} model includes explicitly the
isospin degrees of freedom through different proton and neutron initial
distributions as well as their different mean-field potentials and two-body
collisions in subsequent dynamic evolutions. For nucleon-nucleon cross
sections, we use as default the experimental values in free space. To study
the effects due to the isospin-dependence of in-medium nucleon-nucleon cross
sections $\sigma _{\mathrm{medium}}$, we also use a parameterization
obtained from the Dirac-Brueckner approach based on the Bonn A potential %
\cite{lgq9394}. For the experimental free-space cross sections, the
neutron-proton cross section is about a factor of $3$ larger than the
neutron-neutron or proton-proton cross section. On the other hand, the
in-medium nucleon-nucleon cross sections used here have smaller values and
weaker isospin dependence than $\sigma _{\exp }$ but strong density
dependence. For the isoscalar potential, we use as default the Skyrme
potential with an incompressibility $K_{0}=380$ MeV. This potential has been
shown to reproduce the transverse flow data from heavy ion collisions
equally well as a momentum-dependent soft potential with $K_{0}=210$ MeV %
\cite{pan93,zhang93}.

The \textrm{IBUU} model is solved with the test particle method \cite{buu}.
Although the mean-field potential is evaluated with test particles, only
collisions among nucleons in each event are allowed. Light cluster
production from coalescence of nucleons is treated similar as
nucleon-nucleon collisions, i.e., only nucleons in the same event are
allowed to coalesce to light clusters. Results presented in the following
are obtained with $20,000$ events using $100$ test particles for a physical
nucleon. In the present study, nucleons are considered as being frozen out
when their local densities are less than $\rho _{0}/8$ and subsequent
interactions do not cause their recapture into regions of higher density.
Other emission criteria, such as taking the nucleon emission time as its
last collision time in the \textrm{IBUU} model, do not change our
conclusions. All calculated results presented in the following are in the
center-of-mass system unless stated otherwise.

The \textrm{EOS} or the energy per particle of an asymmetric nuclear matter
with density $\rho $ and isospin asymmetry $\delta =(\rho _{n}-\rho
_{p})/\rho $, where $\rho _{n}$ and $\rho _{p}$ are, respectively, its
neutron and proton densities, is usually approximated by a parabolic law %
\cite{li98}, i.e., 
\begin{equation}
E(\rho ,\delta )=E(\rho ,0)+E_{\mathrm{sym}}(\rho )\delta ^{2},
\end{equation}%
where $E(\rho ,0)$ is the energy per particle of symmetric nuclear matter
and $E_{\mathrm{sym}}(\rho )$ is the nuclear symmetry energy. For the
density dependence of nuclear symmetry energy, we adopt the parameterization
used in Ref. \cite{hei00} for studying the properties of neutron stars,
i.e., 
\begin{equation}
E_{\mathrm{sym}}(\rho )=E_{\mathrm{sym}}(\rho _{0})\cdot u^{\gamma },
\end{equation}%
where $u\equiv \rho /\rho _{0}$ is the reduced density and $E_{\mathrm{sym}%
}(\rho _{0})=30$ \textrm{MeV} is the symmetry energy at normal nuclear
matter density. In the present study, we consider the two cases of $\gamma
=0.5$ (soft) and $2$ (stiff) to explore the large range of $E_{\mathrm{sym}%
}(\rho )$ predicted by many-body theories \cite{bom}.

\begin{figure}[th]
\includegraphics[scale=1.1]{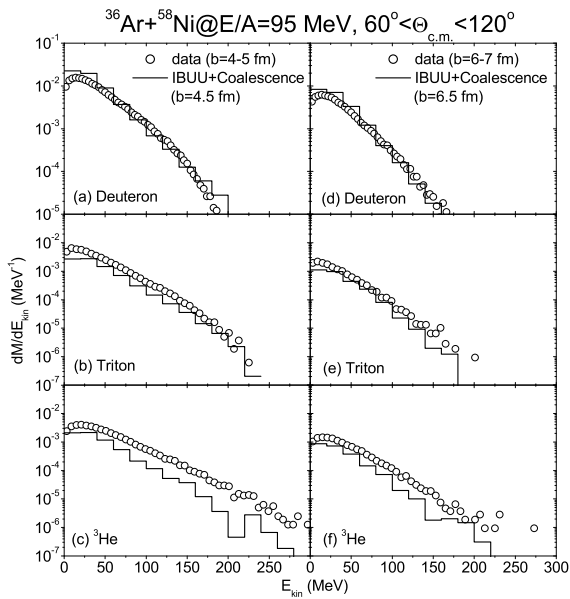}
\caption{{\protect\small Experimental (open circles) and theoretical
(histogram) energy spectra of deuterons ((a) and (d)), tritons ((b) and
(e)), and }$^{3}${\protect\small He ((c) and (f)) from }$^{36}$%
{\protect\small Ar + }$^{58}${\protect\small Ni collisions at }$E/A=95$%
{\protect\small \ MeV for impact parameter }$b=4.5${\protect\small \ (left
panels) and }$6.5${\protect\small \ (right panels) fm and }$\Theta _{\text{%
c.m.}}=60^{\circ }$-$120^{\circ }${\protect\small .}}
\label{ArNi95}
\end{figure}

Before investigating isospin effects on light cluster production from
heavy-ion collisions induced by radioactive nuclei, we first compare
theoretical results obtained from the coalescence model using nucleon phases
space distribution functions from the IBUU model with existing experimental
data from the collisions $^{36}$Ar + $^{58}$Ni at $E/A=95$ \textrm{MeV} at
two impact parameters $b=4.5$ and $6.5$ \textrm{fm}. In the \textrm{IBUU}
model simulations, we use the default parameters, i.e., the hard \textrm{EOS}
with $K_{0}=380$ \textrm{MeV}, the experimental free-space cross sections $%
\sigma _{\exp }$, and the soft symmetry potential. However, the theoretical
results are not corrected by experimental filters, and this may be partially
responsible for their deviations from the experimental data. In Fig. \ref%
{ArNi95}, the energy spectra of deuteron, triton, and $^{3}$He at
center-of-mass angles $\Theta _{\text{c.m.}}=60^{\circ }$-$120^{\circ }$ are
shown in Fig. \ref{ArNi95} by solid curves for the coalescence model and by
open circles for the experimental data \cite{Indra00}. We see that the
theoretical results reproduce satisfactorily the experimental deuteron
energy spectra for both mid-central ($b=4.5$ \textrm{fm}) and mid-peripheral
($b=6.5$ \textrm{fm}) collisions except that they overestimate the yield at
low kinetic energies. For the triton kinetic energy spectra, theoretical
results also agree reasonably with experimental data at both impact
parameters except at low kinetic energies, where they are below the measured
triton yield. For $^{3}$He, both the yield and slope parameter of their
kinetic energy spectra from theoretical calculations are somewhat smaller
than experimental data. The flatter or harder energy spectra of $^{3}$He
than that of deuterons and tritons, known as the ``$^{3}$He puzzle'', has
been seen in other experiments as well \cite%
{Posk71,Lisa95,Poggi95,Marie97,Viola99,Hagel00}.

For a more quantitative comparison, we show in Table \ref{ibuuexp} the
predicted multiplicities and average kinetic energies per nucleon of
deuteron, triton, and $^{3}$He in center-of-mass angles $\Theta _{\text{c.m.}%
}=60^{\circ }$-$90^{\circ }$ together with corresponding experimental data.
It is seen that the deuteron multiplicity from our model is about $40\%$
above the experimental data, while those for triton and $^3$He are about $%
50\%$ below the experimental data. As to the average kinetic energy per
nucleon, our model describes reasonably those of deuteron and triton but
underestimates that of $^{3}$He. Failure of our model for $^{3}$He
production may be related to the neglect in the coalescence model of its
binding energy, which is larger than those of deuteron and triton. Also, the
harmonic oscillator wave functions used for $^{3}$He may not be as good an
approximation for tritons. Furthermore, in the present dynamic coalescence
model, we have neglected the final-state long range Coulomb potential acting
on protons after they freeze out, which would increase the kinetic energy of
produced clusters, particularly $^{3}$He that has a larger charge to mass
ratio than deuterons and tritons. In addition, the momentum-dependence of
the isoscalar nuclear potential or symmetry potential, which is not
considered in the present work, may change the particle energy spectra.
Although these corrections need to be included for a more quantitative
study, the symmetry energy effect on cluster production shown below is not
expected to be appreciably affected.

%\vspace{-0.58cm}

\begin{table}[tbp]
\caption{{\protect\small Calculated multiplicities and average kinetic
energies per nucleon (MeV/nucleon) of deuteron, triton, and }$^{3}$%
{\protect\small He together with the corresponding experimental data %
\protect\cite{Indra00}.}}%
\begin{tabular}{cccccc}
\hline\hline
$b$ $($fm$)$ & \quad Particle\quad  & $\quad M_{60^{\circ }-90^{\circ
}}^{\exp }$ & $M_{60^{\circ }-90^{\circ }}^{\text{cal}}$ & $E_{60^{\circ
}-90^{\circ }}^{\exp }$ & $E_{60^{\circ }-90^{\circ }}^{\text{cal}}$ \\ 
\hline
& d & 0.814 & 1.15 & 18.2 & 15.7 \\ 
$4$-$5$ & t & 0.308 & 0.163 & 11.6 & 11.9 \\ 
& $^{3}$He & 0.251 & 0.128 & 14.7 & 11.9 \\ \hline
& d & 0.291 & 0.410 & 15.9 & 14.7 \\ 
$6$-$7$ & t & 0.090 & 0.057 & 9.8 & 10.3 \\ 
& $^{3}$He & 0.077 & 0.044 & 12.0 & 10.4 \\ \hline\hline
\end{tabular}%
\label{ibuuexp}
\end{table}

\section{Isospin effects on light cluster production}

\label{results}

To study isospin effects on cluster production in heavy ion collisions, we
consider mainly the reaction system of $^{52}$Ca + $^{48}$Ca, which has an
isospin asymmetry $\delta =0.2$ and can be studied at future Rare Isotope
Accelerator Facility. Another heavier reaction system of $^{132}$Sn + $^{124}
$Sn, which has a similar isospin asymmetry, will also be investigated to see
how the isospin effect changes with the mass of the reaction system. In
particular, we have considered the collisions $^{52}$Ca+$^{48}$Ca and also $%
^{132}$Sn+$^{124}$Sn at an incident energy of 80 MeV/nucleon. We choose
these reaction systems and incident energy as effects due to feedback from
heavy fragment evaporation and feed-down from highly excited fragments to
light cluster production, which cannot be described by our model, are
expected to be less important than in collisions between heavier nuclei at
lower energies.

\subsection{Yields and energy spectra of light clusters}

\begin{figure}[th]
\includegraphics[scale=1.1]{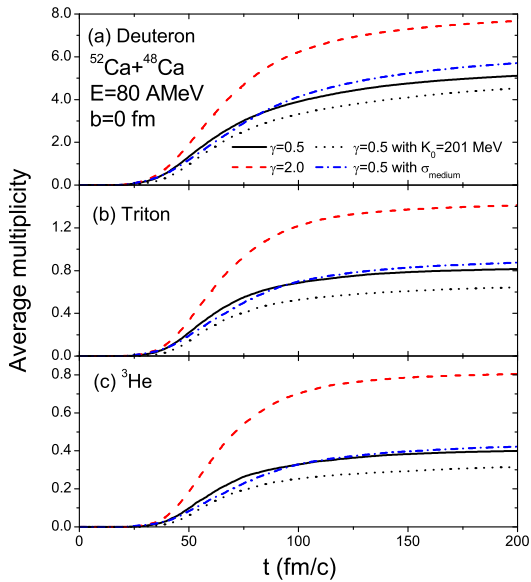}
\caption{{\protect\small Time evolution of the average multiplicities of (a)
deuterons, (b) tritons, and (c) }$^{3}${\protect\small He from central
collisions of }$^{52}${\protect\small Ca+}$^{48}${\protect\small Ca at }$%
E=80 ${\protect\small \ MeV/nucleon by using the soft (solid curves) or
stiff (dashed curves) symmetry energy with a stiff nuclear compressibility }$%
K_{0}=380${\protect\small \ MeV and free nucleon-nucleon cross sections.
Results using soft symmetry energy and free nucleon-nucleon cross sections
but }$K_{0}=201${\protect\small \ MeV are shown by dotted curves, while
those from soft symmetry energy and }$K_{0}=380${\protect\small \ MeV but
in-medium nucleon-nucleon cross sections are given by dash-dotted curves.}}
\label{Mult}
\end{figure}

The yields and energy spectra of light clusters from heavy ion collisions
are important experimental observables that can provide useful information
about particle production mechanism and reaction dynamics. Shown in Figs. %
\ref{Mult} (a), (b) and (c) is the time evolution of the average
multiplicities of deuteron, triton, and $^{3}$He from central collisions of $%
^{52}$Ca + $^{48}$Ca at $E=80$ MeV/nucleon by using the soft (solid curve)
or stiff (dashed curve) symmetry energy in the IBUU model. It is seen that
production of these light clusters from a neutron-rich reaction system is
sensitive to the density dependence of nuclear symmetry energy. The final
multiplicities of deuteron, triton, and $^{3}$He for the stiff symmetry
energy are larger than those for the soft symmetry energy by $51\%$, $73\%$,
and $100\%$, respectively. This is due to the fact that the stiff symmetry
energy induces a stronger pressure in the reaction system and thus causes an
earlier emission of neutrons and protons than in the case of the soft
symmetry energy \cite{CorLong}, leading to a stronger correlations among
nucleons. Furthermore, the soft symmetry energy, which gives a more
repulsive symmetry potential for neutrons and more attractive one for
protons in low density region ($\leq \rho _{0}$) than those from the stiff
symmetry energy, generates a larger phase-space separation between neutrons
and protons at freeze out and thus a weaker correlations among nucleons. The
larger sensitivity of the multiplicity of $^{3}$He to the nuclear symmetry
energy than those of triton and deuteron seen in Fig. \ref{Mult} reflects
the fact that the effect of symmetry energy is stronger on protons than
neutrons with lower momenta \cite{CorLong}.

We have also studied the effects of using different nuclear
incompressibilities and nucleon-nucleon cross sections. As shown in Fig. \ref%
{Mult}, changing the incompressibility from $K_{0}=380$ to $201$ MeV (dotted
curves) or using $\sigma _{\mathrm{medium}}$ instead of $\sigma _{\mathrm{exp%
}}$ (dash-dotted curves) only leads to a small change in the yields of light
clusters. This implies that the relative space-time structure of neutrons
and protons at freeze out is not sensitive to the EOS of symmetric nuclear
matter and the medium dependence of nucleon-nucleon cross sections. As
explained in Ref.\cite{CorShort}, this is due to the fact that the two
symmetric-matter EOS's create a similar pressure effect on particle
emissions as they lead to different density evolutions with the soft one
giving a larger maximum density.

\begin{figure}[th]
\includegraphics[scale=1.1]{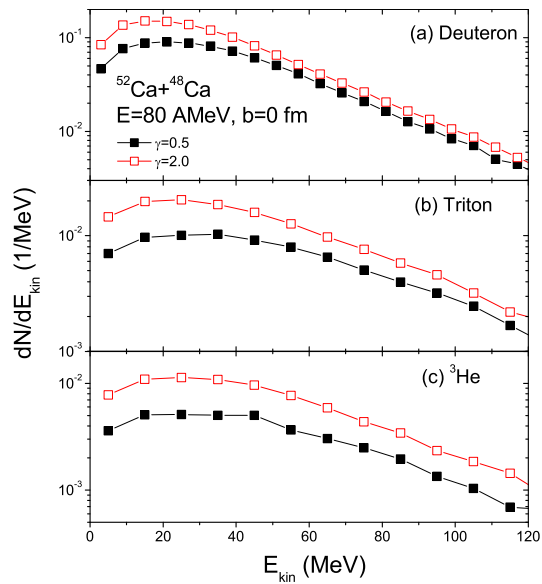}
\caption{{\protect\small Kinetic energy spectra in center-of-mass system for
(a) deuterons, (b) tritons, and (c) }$^{3}${\protect\small He from central
collisions of }$^{52}${\protect\small Ca+}$^{48}${\protect\small Ca at }$%
E=80 ${\protect\small \ MeV/nucleon by using the soft (solid squares) or
stiff (open squares) symmetry energy with a stiff nuclear compressibility }$%
K_{0}=380${\protect\small \ MeV and free nucleon-nucleon cross sections.}}
\label{kinetic}
\end{figure}

The kinetic energy spectra in the center-of-mass system for deuterons,
tritons, and $^{3}$He's are shown in Fig. \ref{kinetic} for both the soft
(solid squares) and stiff (open squares) symmetry energies. It is seen that
the symmetry energy has a stronger effect on the yield of low energy
clusters than that of high energy ones, although the effect is still
appreciable for high energy clusters. For example, the symmetry energy
effect on the yield of deuteron, triton, and $^{3}$He is about $60\%$, $%
100\% $, and $120\%$, respectively, if their kinetic energies are around $10$
\textrm{MeV}, but is about $30\%$, $40\%$, and $85\%$, respectively, if
their kinetic energies are around $100$ \textrm{MeV}. This follows from the
fact that lower energy clusters are emitted later in time when the size of
nucleon emission source is relatively independent of nuclear symmetry
energy, leading thus to similar probabilities for nucleons to form light
clusters. Since there are more low energy nucleons for a stiffer symmetry
energy, more light clusters are produced. On the other hand, higher energy
nucleons are emitted earlier when the size of emission source is more
sensitive to the symmetry energy, with a smaller size for a stiffer symmetry
energy. The probability for light cluster formation is thus larger for a
stiffer symmetry energy. This effect is, however, reduced by the smaller
number of high energy nucleons if the symmetry energy is stiffer. As a
result, production of high energy light clusters is less sensitive to the
stiffness of symmetry energy. This is different from that seen in the
correlation functions between two nucleons with low relative momentum, where
the symmetry energy effect is larger for nucleon pairs with higher kinetic
energies \cite{CorShort}, as they are affected only by the size of emission
source but not by the number of emitted nucleon pairs.

\subsection{The isobaric yield ratio t/$^{3}$He}

\begin{figure}[th]
\includegraphics[scale=0.85]{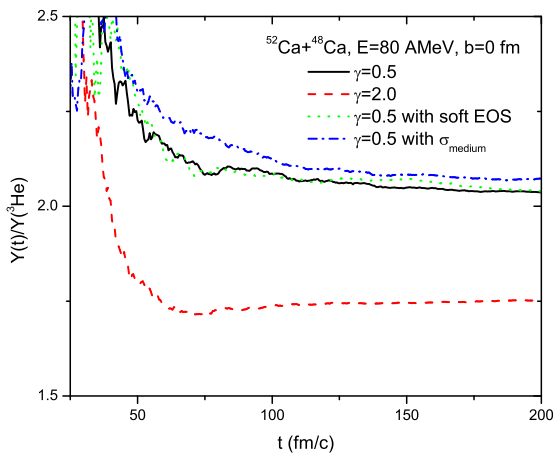}
\caption{{\protect\small Time evolution of the isobaric yield ratio t/}$^{3}$%
{\protect\small He from central collisions of }$^{52}${\protect\small Ca + }$%
^{48}${\protect\small Ca at }$E=80${\protect\small \ MeV/nucleon using
different symmetry energies, nuclear incompressibilities, and \textsl{N}-%
\textsl{N} cross sections.}}
\label{RtHe3t}
\end{figure}

To minimize the model dependence and also other effects, such as the
feedback from heavy fragment evaporation and the feed-down from produced
excited states of triton and $^{3}$He, it is of interest to study the
isobaric yield ratio t/$^{3}$He. In Fig. \ref{RtHe3t}, we show the time
evolution of the isobaric yield ratio t/$^{3}$He from central collisions of $%
^{52}$Ca + $^{48}$Ca at $E=80$ \textrm{MeV/nucleon } for different symmetry
energies. It is seen that the stiff symmetry energy gives a final isobaric
yield ratio t/$^{3}$He that is about $17\%$ less than that for the soft
symmetry energy. This is due to the fact that a soft symmetry energy
increases neutron emission and reduces proton emission while a stiff
symmetry energy enhances the emission of both neutrons and protons. As a
result, the t/$^{3}$He ratio is enhanced using a soft symmetry energy but is
reduced using a stiff symmetry energy. It is further seen that the effect of
symmetry energy on the t/$^{3}$He ratio is more pronounced at earlier stage
of reactions, e.g., at $70$ \textrm{fm/c} the t/$^{3}$He ratio is increased
by $23\%$ from the stiff to the soft symmetry energy.

We have also studied the effect of nuclear incompressibility and \textsl{N}-%
\textsl{N} cross sections on the t/$^{3}$He ratio. As shown in Fig.\ref%
{RtHe3t}, changing nuclear incompressibility from $K_{0}=380$ to $201$ 
\textrm{MeV} or using $\sigma _{\text{\textrm{medium}}}$ instead of $\sigma
_{\exp }$ increases the ratio t/$^{3}$He by less than $4\%$. These results
imply again that the relative space-time structure of neutrons and protons
at freeze-out is not sensitive to the EOS of symmetric nuclear matter and
the medium dependence of nucleon-nucleon cross sections.

\begin{figure}[th]
\includegraphics[scale=0.85]{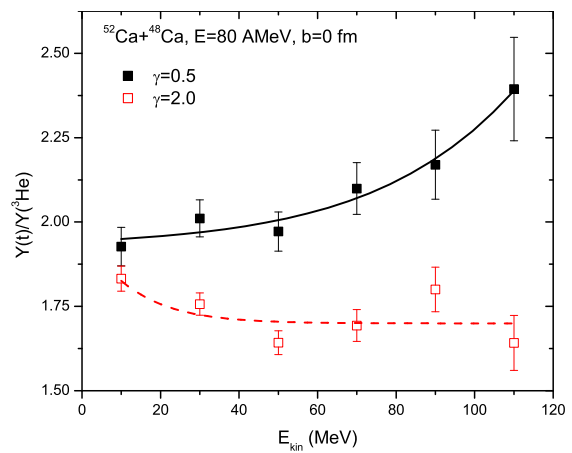}
\caption{{\protect\small The isobaric ratio t/}$^{3}${\protect\small He as a
function of cluster kinetic energy in the center-of-mass system by using the
soft or stiff symmetry energy. The lines are drawn to guide the eye.}}
\label{RtHe3E}
\end{figure}

To further investigate the symmetry energy effect, we show in Fig. \ref%
{RtHe3E} the t/$^{3}$He ratio with statistical errors as a function of
cluster kinetic energy in the center-of-mass system for the soft (solid
squares) and stiff (open squares) symmetry energies. It is seen that the
ratio t/$^{3}$He obtained with different symmetry energies has very
different energy dependence. While the t/$^{3}$He ratio increases with
kinetic energy for the soft symmetry energy, it decreases with kinetic
energy for the stiff symmetry energy. The symmetry energy thus affects more
strongly the ratio of high kinetic energy triton and $^{3}$He. For both soft
and stiff symmetry energies, the ratio t/$^{3}$He is larger than the neutron
to proton ratio of the whole reaction system, i.e., \textsl{N/Z}$=1.5$. This
is in agreement with available experimental results and also with
statistical model simulations in other reaction systems and incident
energies \cite{Naga81,Cibor00,Hagel00,Sobotka01,Vesel01,Chomaz99,Gupta01}.
It is interesting to note that although the yield of lower energy triton and 
$^{3}$He is more sensitive to symmetry energy than higher energy ones, as
shown in Fig. \ref{kinetic}, their ratio at higher energy is affected more
by the symmetry energy. Moreover, the energy dependence of the ratio t/$^{3}$%
He is insensitive to the \textrm{EOS} of symmetric nuclear matter and the
medium dependence of nucleon-nucleon cross sections. Therefore, the
pre-equilibrium triton to $^{3}$He ratio is a sensitive probe to the density
dependence of nuclear symmetry energy.

\subsection{Centrality, incident energy, and reaction system dependence}

\begin{figure}[th]
\includegraphics[scale=1.25]{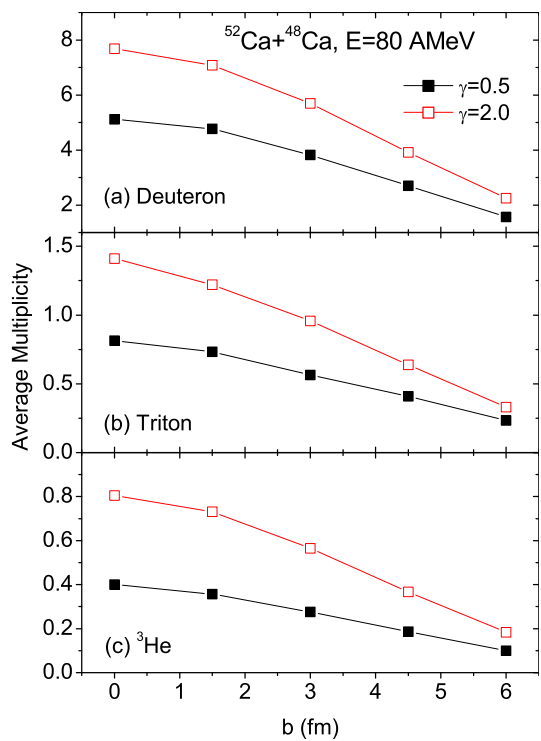}
\caption{{\protect\small Impact parameter dependence of the yield for (a)
deuterons, (b) tritons, and (c) $^3$He, by using $K_{0}=380$ MeV with the
soft (filled squares) or stiff (open squares) symmetry energy.}}
\label{Bdep}
\end{figure}

Light cluster production in heavy ion collisions also depends on the impact
parameter, incident energy, and the mass of the reaction system. Fig. \ref%
{Bdep} shows the impact parameter dependence of light cluster yield by using 
$K_{0}=380$ \textrm{MeV} with the soft (filled squares) and stiff (open
squares) symmetry energies. It is seen that the symmetry energy effect on
the yield of light clusters decreases slightly with increasing impact
parameter. This simply reflects the fact that nuclear compression becomes
weaker when the impact parameter increases and the symmetry energy effect
mainly comes from the difference between the soft and stiff symmetry
energies at low densities. Also, we find that the light cluster yield
decreases with increasing impact parameter as a result of smaller number of
free nucleons in collisions at larger impact parameters.

\begin{figure}[th]
\includegraphics[scale=1.25]{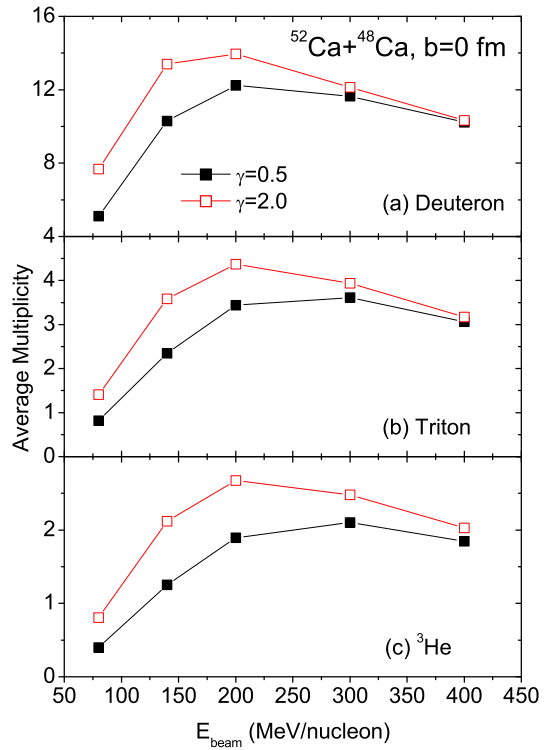}
\caption{{\protect\small Dependence of the yield for (a) deuterons, (b)
tritons, and (c) }$^{3}${\protect\small He on beam energy. The results are
obtained using }$K_{0}=380${\protect\small \ MeV with the soft (filled
squares) or stiff (open squares) symmetry energy.}}
\label{Edep}
\end{figure}

Dependence of the light cluster yield on the incident energy of heavy ion
collisions is shown in Fig. \ref{Edep}. The results are obtained using $%
K_{0}=380$ \textrm{MeV} with the soft (filled squares) or stiff (open
squares) symmetry energy. These results show that for the light clusters
considered here their yields first increase and then decrease with
increasing incident energy. The initial increase is due to increasing
compression of nuclear matter with increasing incident energy. As a result,
nucleons are emitted earlier and the source size for cluster production is
smaller, leading thus to a larger probability for cluster production. On the
other hand, with further increase in incident energy, the phase-space volume
occupied by nucleons becomes larger, which reduces the formation probability
of light clusters. These features have been observed in previous experiments %
\cite{Naga81}. We further see that the symmetry energy effect on light
cluster production becomes weaker at higher incident energies. This is due
to the fact that at higher incident energies the symmetry energy effect
mainly comes from early compression stage of heavy ion collisions when the
isoscalar part of nuclear equation of state dominates. Therefore, the
colliding system expands rapidly at high incident energies, and the low
density behavior of symmetry potential thus plays a less important role. On
the other hand, the initial compression is relatively weak at lower incident
energies, and the behavior of symmetry energy at low densities becomes
relevant.

\begin{figure}[th]
\includegraphics[scale=1.15]{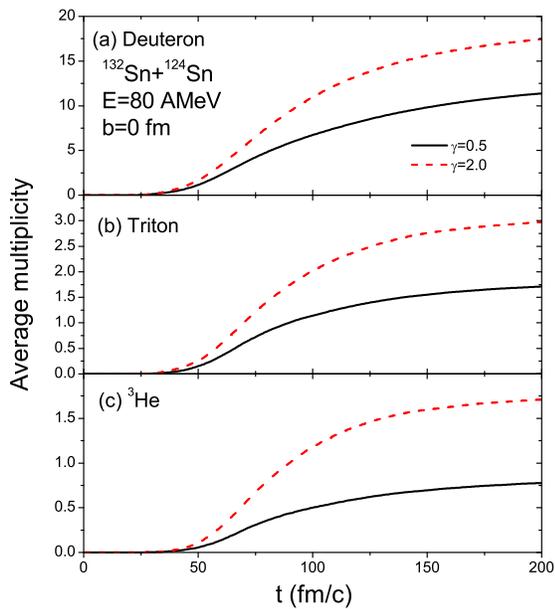}
\caption{{\protect\small Time evolution of the average multiplicity of (a)
deuteron, (b) triton, and (c) }$^{3}${\protect\small He from central
collisions of }$^{132}${\protect\small Sn + }$^{124}${\protect\small Sn at }$%
80${\protect\small \ MeV/nucleon by using }$K_{0}=380${\protect\small \ MeV
with the soft (solid curve) or stiff (dashed curve) symmetry energy.}}
\label{MultSn}
\end{figure}

To see how the yield of light clusters changes in different reaction
systems, we shown in Fig. \ref{MultSn} the time evolution of the average
multiplicities of deuteron, triton, and $^{3}$He from central collisions of
the reaction system $^{132}$Sn + $^{124}$Sn at $80$ \textrm{MeV/nucleon} by
using $K_{0}=380$ \textrm{MeV} with the soft (solid line) or stiff (dashed
line) symmetry energy. This reaction system has a similar isospin asymmetry,
i.e., $\delta =0.22$, as the reaction system $^{52}$Ca + $^{48}$Ca studied
in the above, and can also be studied in future RIA. It is seen that the
light cluster yield in $^{132}$Sn + $^{124}$Sn reactions is larger than that
in $^{52}$Ca + $^{48}$Ca reactions shown in Fig. \ref{Mult} as a result of
more free neutrons in the heavier reaction system. The symmetry energy
effect is slightly stronger in $^{132}$Sn + $^{124}$Sn reactions than in $%
^{52}$Ca + $^{48}$Ca. Explicitly, the final multiplicities of deuteron,
triton, and $^{3}$He for the stiff symmetry energy are larger than those for
the soft symmetry energy by $53\%$, $74\%$, and $120\%$, respectively.

\section{Correlation functions of light clusters}

\label{correlation}

The particle emission function, which is important for understanding the
reaction dynamics in heavy ion collisions, can be extracted from
two-particle correlation functions; see, e.g., Refs. \cite%
{Boal90,bauer,ardo97,Wied99} for reviews. Most extensively studied one, both
experimentally and theoretically, has been the two-proton correlation
function \cite{gong90,gong91,kunde93,handzy95,Verde02,Verde03}. Recently,
data on two-neutron, neutron-proton correlation functions have also become
available. The neutron-proton correlation function is especially useful as
it is free of correlations due to wave-function anti-symmetrization and
Coulomb interactions. Indeed, Ghetti \textit{et al.} have deduced from the
measured neutron-proton correlation function the time sequence of neutron
and proton emissions \cite{Ghetti00,Ghetti01}. The fragment-fragment
correlation functions have also provided useful information about the
particle production mechanism and reaction dynamics in heavy ion collisions
at intermediate energies \cite%
{Chitwood85,Poch85,Kryger90,Zhu91,gong93,Eraz94,Hamilton96,Kotte99,Gourio00,Ghetti03}%
.

\subsection{Average emission times of light clusters}

\begin{figure}[th]
\includegraphics[scale=1.1]{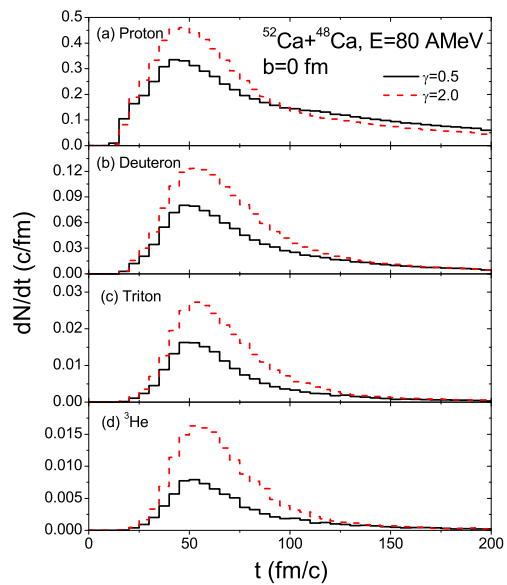}
\caption{{\protect\small Emission time distributions for (a) protons, (b)
deuterons, (c) tritons, and (d) }$^{3}${\protect\small He by using }$%
K_{0}=380${\protect\small \ MeV with the soft (solid line) or stiff (dashed
line) symmetry energy.}}
\label{emProb}
\end{figure}

The emission times of different particles in heavy ion collisions are
relevant for understanding both the collision dynamics and the mechanism of
particle production. In heavy ion collisions at intermediate energies,
nucleon emissions are mainly governed by the pressure of the excited nuclear
matter during the initial stage of collisions \cite{lis,pawel}. Since the
stiff symmetry energy gives a larger pressure than that due to the soft
symmetry energy \cite{CorLong}, it leads to an earlier emission of neutrons
and protons. To see the symmetry energy effect on light cluster emissions,
we show in Figs. \ref{emProb} (a), (b), (c) and (d) the emission time
distributions for protons, deuterons, tritons, and $^{3}$He, respectively,
by using $K_{0}=380$ \textrm{MeV} with the soft (solid line) or stiff
(dashed line) symmetry energy. One sees that the emission time distributions
of light clusters are different from that of protons. While the proton
emission time peaks earlier at $\sim 45$ \textrm{fm/c} and lasts longer, the
emission time of light clusters peaks later at $\sim 55$ \textrm{fm/c} and
has a shorter duration. Light cluster production thus occurs mainly from the
pre-equilibrium stage of heavy ion collisions as in previous BUU model
simulations \cite{gong93}. As to the symmetry energy effect on particle
emissions, it is seen from Fig. \ref{emProb} that the particle emission
times are earlier for the stiff symmetry energy than for the soft symmetry
energy as the former gives a larger initial pressure.

\begin{figure}[th]
\includegraphics[scale=1.15]{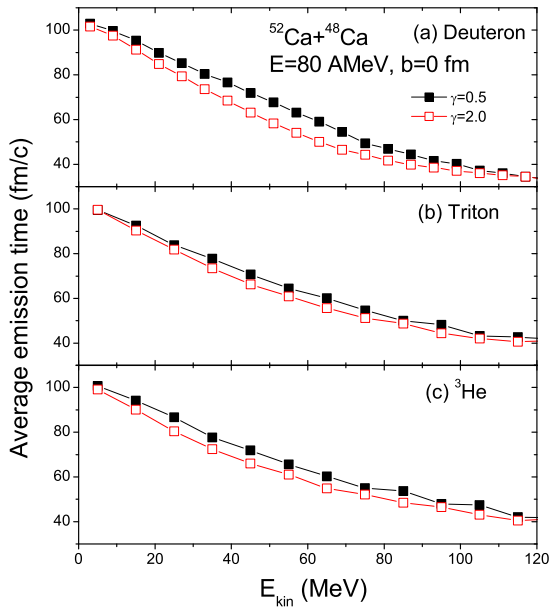}
\caption{{\protect\small Average emission times of (a) deuterons, (b)
tritons, and (c) }$^{3}${\protect\small He as functions of their c.m.
kinetic energies by using }$K_{0}=380${\protect\small \ MeV with the soft
(filled squares) or stiff symmetry energy (open squares).}}
\label{emTimeE}
\end{figure}

Particles emitted in earlier stage of heavy ion collisions usually have
higher energy than those emitted during later stage of the reaction. It is
thus of interest to study the correlation between the emission time of a
particle and its energy. Shown in Fig. \ref{emTimeE} are the average
emission times of deuterons, tritons, and $^{3}$He as functions of their
c.m. kinetic energies by using $K_{0}=380$ \textrm{MeV} with the soft or
stiff symmetry energy. It is seen that the average emission time of a
particle with a given c.m. kinetic energy is earlier for the stiff symmetry
energy (open squares) than for the soft one (filled squares). These features
are consistent with results shown in Fig. \ref{emProb} on the particle
emission time distributions. It is also worth noting that the emission time
of a particle also depends on its energy, i.e., particles with higher
kinetic energies are emitted earlier than those with lower kinetic energies.

\begin{figure}[th]
\includegraphics[scale=1.15]{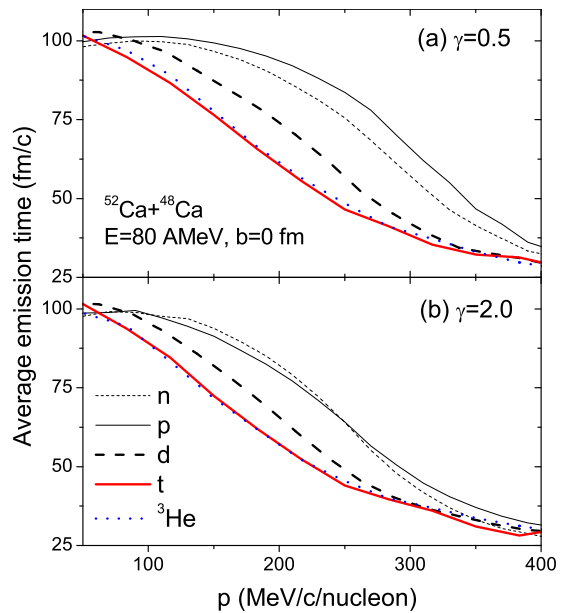}
\caption{{\protect\small Average emission times of neutrons, protons,
deuterons, tritons, and }$^{3}${\protect\small He as functions of their
momenta per nucleon by using }$K_{0}=380${\protect\small \ MeV with the soft
(upper panel) or stiff (lower panel) symmetry energy.}}
\label{TorderP}
\end{figure}

To see the effect of symmetry energy on the emission sequence of different
particles, we show in Fig. \ref{TorderP} the average emission times of
neutrons, protons, deuterons, tritons, and $^{3}$He as functions of their
momenta per nucleon by using $K_{0}=380$ \textrm{MeV} with the soft (upper
panel) or stiff (lower panel) symmetry energy. It is seen that ordering of
the average emission times for neutrons and protons depends on the stiffness
of symmetry energy, with a significant delay of proton emission in the case
of soft symmetry energy. The ordering of average emission times for light
clusters depends further on their masses, with heavier clusters emitted
earlier than lighter ones at the same momentum per nucleon, which is
expected from the coalescence model. We note that although the density
dependence of nuclear symmetry energy influences significantly the ordering
of neutron and proton average emission times, it has only a weak effect on
that for tritons and $^{3}$He. Experimentally, the particle emission
sequence can be studied from the correlation functions of non-identical
particles using various energy cuts \cite%
{Gelder95,Ghetti00,Ghetti01,Ghetti03}.

\subsection{Correlation functions of deuteron-deuteron, triton-triton, and $%
^{3}$He-$^{3}$He}

In standard Koonin-Pratt formalism \cite{koonin77,pratt1,pratt2}, the
two-particle correlation function is obtained by convoluting the emission
function $g(\mathbf{p},x)$, i.e., the probability for emitting a particle
with momentum $\mathbf{p}$ from the space-time point $x=(\mathbf{r},t)$,
with the relative wave function of the two interacting particles, i.e., 
\begin{equation}
C(\mathbf{P},\mathbf{q})=\frac{\int d^{4}x_{1}d^{4}x_{2}g(\mathbf{P}%
/2,x_{1})g(\mathbf{P}/2,x_{2})\left| \phi (\mathbf{q},\mathbf{r})\right| ^{2}%
}{\int d^{4}x_{1}g(\mathbf{P}/2,x_{1})\int d^{4}x_{2}g(\mathbf{P}/2,x_{2})}.
\label{Eq1}
\end{equation}%
In the above, $\mathbf{P(=\mathbf{p}_{1}+\mathbf{p}_{2})}$ and $\mathbf{q(=}%
\frac{1}{2}(\mathbf{\mathbf{p}_{1}-\mathbf{p}_{2}))}$ are, respectively, the
total and relative momenta of the particle pair; and $\phi (\mathbf{q},%
\mathbf{r})$ is the relative two-particle wave function with $\mathbf{r}$
being their relative position, i.e., $\mathbf{r=(r}_{2}\mathbf{-r}_{1}%
\mathbf{)-}$ $\frac{1}{2}(\mathbf{\mathbf{v}_{1}+\mathbf{v}_{2})(}t_{2}-t_{1}%
\mathbf{)}$. This approach has been shown to be very useful in studying the
reaction dynamics of intermediate energy heavy-ion collisions \cite{bauer}.
In the present paper, we use the Koonin-Pratt method to determine the
particle-particle correlation functions in order to study the effect due to
the density dependence of nuclear symmetry energy on the spatial and
temporal structure of the emission source for light clusters in intermediate
energy heavy ion collisions.

\begin{figure}[th]
\includegraphics[scale=0.8]{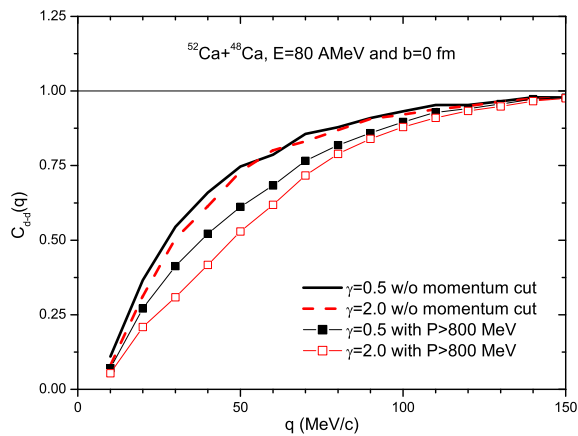}
\caption{{\protect\small Two-deuteron correlation functions from central
collisions of }$^{52}${\protect\small Ca + }$^{48}${\protect\small Ca at }$%
E=80${\protect\small \ MeV/nucleon. Results without momentum cut are shown
by solid and dashed curves for the soft and stiff symmetry energies,
respectively. Curves with filled and open squares are corresponding ones for
deuterons with total momentum }$P>800${\protect\small \ MeV/c.}}
\label{CorDD}
\end{figure}

Using the program Correlation After Burner \cite{hbt}, which takes into
account final-state particle-particle interactions, we have evaluated
two-cluster correlation functions from the emission function given by the
IBUU model. We first show in Fig. \ref{CorDD} the two-deuteron correlation
functions from central collisions of $^{52}$Ca + $^{48}$Ca at $E=80$ \textrm{%
MeV/nucleon}. Solid and dashed curves are results without momentum cut using
the soft and stiff symmetry energies, respectively. The results for deuteron
pairs with total momenta $P>800$ \textrm{MeV/c} are shown, respectively, by
curves with filled and open squares for the soft and stiff symmetry
energies. In obtaining these results, the final-state strong interaction
between two deuterons includes only the relative $s$-wave interaction, which
we use the one in Ref. \cite{Boal90}\textit{\ }that fits the
deuteron-deuteron elastic scattering phase-shift by Woods-Saxon potential.
Because of the repulsive $s$-wave nuclear potential and Coulomb potential,
two deuterons are anti-correlated as shown in Fig. \ref{CorDD}, and the
anti-correlation is stronger for deuteron pairs with total momenta $P>800$ 
\textrm{MeV/c}. The latter implies that deuterons with higher momenta are
emitted earlier and thus have a smaller emission source size. Although the
correlation function for all deuterons depends weakly on the stiffness of
nuclear symmetry energy, that for deuterons with total momenta $P>800$ 
\textrm{MeV/c} clearly shows a stronger anti-correlation for the stiff than
for the soft symmetry energy. For example, the soft symmetry energy gives a
correlation function at $q=40$ \textrm{MeV/c} that is $20\%$ larger than
that due to the stiff symmetry energy. Therefore, pre-equilibrium deuterons
with higher momenta are emitted earlier or have a smaller source size for
the stiff symmetry energy, consistent with the results shown in Fig. \ref%
{emTimeE} on their average emission times. The correlation function of
deuteron pairs with high total momenta is thus sensitive to the density
dependence of nuclear symmetry energy. 
\begin{figure}[th]
\includegraphics[scale=0.8]{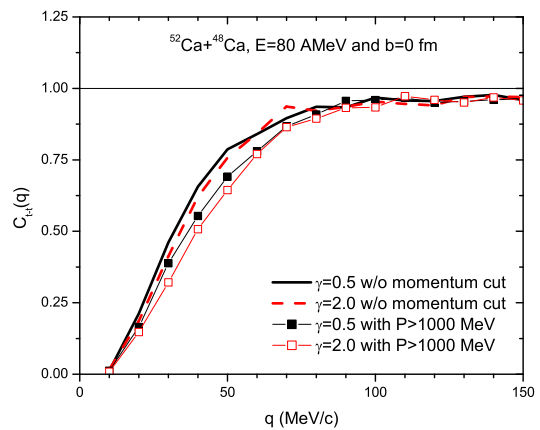}
\caption{{\protect\small Same as Fig. \ref{CorDD} for two-triton correlation
function with and without momentum cut.}}
\label{CorTri}
\end{figure}
\begin{figure}[th]
\includegraphics[scale=0.8]{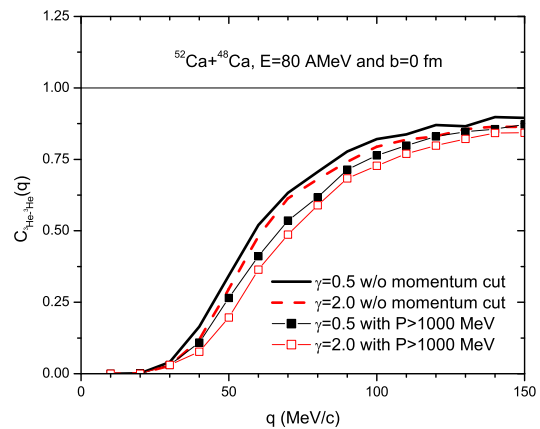}
\caption{{\protect\small Same as Fig. \ref{CorDD} for two-$^{3}$He
correlation function with and without momentum cut.}}
\label{CorHe3}
\end{figure}

In Figs. \ref{CorTri} and \ref{CorHe3}, we show, respectively, the
two-triton and two-$^{3}$He correlation functions with and without momentum
cut. Since the strong interaction potential between two tritons or $^{3}$He
is poorly known, the results obtained here include only the Coulomb
potential in the final-state interaction as in Ref. \cite{Boal90}. It is
seen that the stiff symmetry energy leads to a slightly stronger
anti-correlation of triton or $^{3}$He pairs than that obtained with the
soft symmetry energy, particularly for triton or $^{3}$He pairs with total
momenta $P>1000$ \textrm{MeV/c}. The symmetry energy effect on heavier
clusters is thus weaker than on lighter particles such as nucleons and
deuterons, and this is consistent with the fact that the symmetry energy
effect on cluster emission times becomes weaker with increasing particle
mass as shown in Fig. \ref{emTimeE}.

\section{Summary and outlook}

\label{summary}

Using an isospin-dependent transport model, we have studied light cluster
production via the coalescence model in heavy-ion collisions induced by
neutron-rich nuclei. We find that the density dependence of nuclear symmetry
energy affects significantly the yield of deuterons, tritons, and $^3$He in
these collisions. More light clusters are produced for the stiff symmetry
energy than the soft symmetry energy. This effect is particularly large when
these clusters have lower kinetic energies. Also, the isobaric ratio t/$^{3}$%
He, especially for higher energy tritons and $^{3}$He, shows a strong
sensitivity to the density dependence of nuclear symmetry energy. It is
further found that light cluster production is insensitive to the \textrm{EOS%
} of the isospin-independent part of nuclear equation of state and the
medium dependence of nucleon-nucleon cross sections.

We have also studied the dependence of light cluster production on the
impact parameter and incident energy of heavy ion collisions as well as the
mass of the reaction system. We find that the symmetry energy effect on
light cluster production depends weakly on impact parameter and the mass of
the reaction system. However, with increasing incident energy the symmetry
energy effect on light cluster production becomes weaker.

We have further studied the emission time and sequence as well as the
correlation functions of light clusters. It is found that at same momentum
per nucleon heavier clusters are emitted earlier than lighter ones as
expected from the coalescence model. Since the emission time of light
clusters is earlier for the stiff symmetry energy than for the soft one, a
stronger anti-correlation of light clusters is obtained with the stiff
symmetry energy. However, the symmetry energy effect on the correlation
functions of light clusters becomes weaker with increasing cluster mass.

Isospin effects on cluster production and isotopic ratios in heavy ion
collisions have been previously studied using either the lattice gas model %
\cite{Chomaz99} or a hybrid of \textrm{IBUU} and statistical fragmentation
model \cite{tan01}. These studies are, however, at lower energies than
considered here, where effects due to multifragmentation as a result of
possible gas-liquid phase transition may play an important role. Except
deuterons, both tritons and $^{3}$He are only about one per event in heavy
ion collisions at $80$ \textrm{AMeV}. The number of other clusters such as
the alpha particle is not large either \cite{jacak,borderie}. In this case,
the coalescence model is expected to be a reasonable model for determining
the production of light clusters from heavy ion collisions as demonstrated
in Section \ref{ibuu}. This is particularly so for light clusters with large
kinetic energies as they are not contaminated by contributions from decays
of heavy fragments, which are mainly of low kinetic energies and may not be
negligible in intermediate energy heavy ion collisions. Furthermore, the
effect obtained in present study will be enhanced if other clusters are
emitted earlier during the collisions, leading to a larger isospin asymmetry
in the remaining emission source for light clusters. Studies of light
clusters production and correlations in heavy ion collisions induced by
neutron-rich nuclei thus provide another possible tool for extracting useful
information about the density-dependence of nuclear symmetry energy.

In the present work, we have not included momentum dependence in either the
isoscalar mean-field potential or the symmetry potential. The former may
affect the properties of nuclear emission source as it has been shown that a
momentum-dependent mean-field potential reduces nuclear stopping or
increases nuclear transparency \cite{greco99}. With momentum dependence
included in the nuclear symmetry potential, differences between the neutron
and proton potentials are affected not only by the density dependence of
nuclear symmetry energy but also by the magnitude of proton and neutron
momenta \cite{das,ono02}. Since the momentum-dependent symmetry potential
has been shown to reduce significantly the n/p ratio at higher kinetic
energies \cite{bali03}, it is expected to also affect our results on the
kinetic energy dependence of t/$^3$He ratio. It is thus of interest to study
quantitatively how the results obtained in the present study are modified by
the momentum dependence of nuclear mean-field potentials.

\begin{acknowledgments}
We thank Yu-Gang Ma and Joe Natowitz for discussions on light cluster
production in heavy ion collisions, P. Pawlowski for providing us the
experimental data, and G. Verde for discussions on light cluster correlation
functions. This paper was based on work supported by the U.S. National
Science Foundation under Grant Nos. PHY-0098805 and PHY-0088934 as well as
the Welch Foundation under Grant No. A-1358. LWC is also supported by the
National Natural Science Foundation of China under Grant No. 10105008.
\end{acknowledgments}


\begin{thebibliography}{999}
\bibitem{myers} {\small W.D. Myers and W.J. Swiatecki, Nucl. Phys. \textbf{%
A81}, 1 (1966).}

\bibitem{pomorski} {\small K. Pomorski and J. Dudek, Phys. Rev. C \textbf{67}%
, 044316 (2003); nucl-th/0205011.}

\bibitem{ibook} {\small Isospin Physics in Heavy-Ion Collisions at
Intermediate Energies, Eds. Bao-An Li and W. Udo Schr\"{o}der (Nova Science
Publishers, Inc, New York, 2001).}

\bibitem{oya} {\small K. Oyamatsu, I. Tanihata, Y. Sugahara, K. Sumiyoshi,
and H. Toki, Nucl. Phys. \textbf{A634}, 3 (1998).}

\bibitem{brown} {\small B.A. Brown, Phys. Rev. Lett. \textbf{85}, 5296
(2000).}

\bibitem{hor01} {\small C.J. Horowitz, and J. Piekarewicz, Phys. Rev. Lett 
\textbf{86}, 5647 (2001); Phys Rev. C \textbf{63}, 025501 (2001).}

\bibitem{furn02} {\small R.J. Furnstahl, Nucl. Phys. \textbf{A706}, 85
(2002).}

\bibitem{bethe} {\small H.A. Bethe, Rev. Mod. Phys. \textbf{62}, 801 (1990).}

\bibitem{bom} {\small I. Bombaci, in \cite{ibook}, p.35.}

\bibitem{lat01} {\small J.M. Lattimer and M. Prakash, Astr. Phys. Jour. 
\textbf{550}, 426 (2001).}

\bibitem{npa01} {\small Radioactive Nuclear Beams, a special volume of Nucl.
Phys. \textbf{A693}, (2001), Ed. I. Tanihata.}

\bibitem{li98} {\small B.A. Li, C.M. Ko, and W. Bauer, Int. Jour. Phys. E 
\textbf{7}, 147 (1998).}

\bibitem{ditoro99} {\small M. Di Toro, V. Baran, M. Colonna, G. Fabbri, A.
B. Larionov, S. Maccarone, and S. Scalone, Prog. Part. Nucl. Phys. \textbf{42%
}, 125 (1999).}

\bibitem{li97} {\small B.A. Li, C.M. Ko, and Z.Z. Ren, Phys. Rev. Lett. 
\textbf{78}, 1644 (1997).}

\bibitem{fra1} {\small B.A. Li and C.M. Ko, Nucl. Phys. \textbf{A618}, 498
(1997).}

\bibitem{fra2} {\small V. Baran, M. Colonna, M. Di Toro, and A.B. Larionov,
Nucl. Phys. \textbf{A632}, 287 (1998).}

\bibitem{xu00} {\small H.S. Xu \textit{et al.}, Phys. Rev. Lett. \textbf{85}%
, 716 (2000).}

\bibitem{tan01} {\small W.P. Tan \textit{et al.}, Phys. Rev. C \textbf{64},
051901(R) (2001).}

\bibitem{bar02} {\small V. Baran, M. Colonna, M. Di Toro, V. Greco, and M.
Zielinska-Pfab\'e, and H.H. Wolter, Nucl. Phys. \textbf{A703}, 603 (2002).}

\bibitem{betty} {\small M.B. Tsang \textit{et al.}, Phys. Rev. Lett. \textbf{%
86}, 5023 (2001).}

\bibitem{lis} {\small B.A. Li, A.T. Sustich, and B. Zhang, Phys. Rev. C 
\textbf{64}, 054604 (2001).}

\bibitem{li00} {\small B.A. Li, Phys. Rev. Lett. \textbf{85}, 4221 (2000).}

\bibitem{Greco03} {\small V. Greco, V. Baran, M. Colonna, M. Di Toro, T.
Gaitanos, H.H. Wolter, Phys. Lett. \textbf{B562}, 215 (2003);
nucl-th/0212102.}

\bibitem{li02} {\small B.A. Li, Phys. Rev. Lett. \textbf{88}, 192701 (2002);
Nucl. Phys. A\textbf{708}, 365 (2002).}

\bibitem{CorShort} {\small L.W. Chen, V. Greco, C.M. Ko, and B.A. Li, Phys.
Rev. Lett. \textbf{90}, 162701 (2003); nucl-th/0211002.}

\bibitem{mrow92} {\small S. Mrowczynski, Phys. Lett. \textbf{B277}, 43
(1992).}

\bibitem{ClstShort} {\small L.W. Chen, C.M. Ko, and B.A. Li, Phys. Rev. C 
\textbf{68}, 017601 (2003); nucl-th/0302068.}

\bibitem{Hodgson03} {\small P.E. Hodgson and E. B\u{e}t\'{a}k, Phys. Rep. 
\textbf{374}, 1 (2003); and the references therein.}

\bibitem{Csernai86} {\small \ L.P. Csernai and J.I. Kapusta, Phys. Rep. 
\textbf{131}, 223 (1986); and the references therein.}

\bibitem{Gyu83} {\small M. Gyulassy, K. Frankel, and E.A. Relmer, Nucl.
Phys. A\textbf{402}, 596 (1983).}

\bibitem{Aich87} {\small J. Aichelin, A. Rosenhauer, G. Peilert, H. St\"{o}%
cker, and W. Greiner, Phys. Rev. Lett. \textbf{58}, 1926 (1987).}

\bibitem{Koch90} {\small V. Koch \textit{et al.}, Phys. Lett. B \textbf{241}%
, 174 (1990).}

\bibitem{Mattie95} {\small R. Mattiello \textit{et al.}, Phys. Rev. Lett. 
\textbf{74} 2180 (1995); R. Mattiello \textit{et al.}, Phys. Rev. C \textbf{%
55,} 1443 (1997).}

\bibitem{Nagle96} {\small J. L. Nagle \textit{et al.}, Phys. Rev. C \textbf{%
53}, 367 (1996).}

\bibitem{Indra00} {\small P. Pawlowski \textit{et al.}, Eur. Phys. Jour. A 
\textbf{9,} 371 (2000).}

\bibitem{Pearson63} {\small S.T. Butler and C.A. Pearson, Phys. Rev. \textbf{%
129}, 836 (1963).}

\bibitem{Schwarz63} {\small A. Schwarzschild and C. Zupancic, Phys. Rev. 
\textbf{129}, 854 (1963).}

\bibitem{Gutbrod76} {\small H.H. Gutbrod \textit{et al.}, Phys. Rev. Lett. 
\textbf{37}, 667 (1976).}

\bibitem{Sato81} {\small H. Sato and K. Yazak, Phys. Lett. B \textbf{98},
153 (1981).}

\bibitem{Mekjian78} {\small A.Z. Mekjian, Phys. Rev. C \textbf{17}, 1051
(1978); S. Das Gupta and A.Z. Mekjian, Phys. Rep. \textbf{72}, 131 (1981)}

\bibitem{Heinz99} {\small R. Scheibl and U. Heinz, Phys. Rev. C \textbf{59,}
1585 (1999).}

\bibitem{Polleri99} {\small A. Polleri et al., Nucl. Phys. A\textbf{661},
452c (1999).}

\bibitem{Hodgson71} {\small P. E. Hodgson, \textit{Nuclear Reaction and
Nuclear Structure} (Clarendon, Oxford, 1971), p. 453.}

\bibitem{Ericsson88} {\small T. Ericsson and W. Weise, \textit{Pions and
Nuclei} (Clarendon, Oxford, 1988).}

\bibitem{Shin90} {\small G.R. Shin and J. Rafelski, J. Phys. G \textbf{16},
L187 (1990).}

\bibitem{Dover83} {\small A.T. M. Aerts and C.B. Dover, Phys. Rev. D \textbf{%
28,} 450 (1983).}

\bibitem{chen86} {\small C.R. Chen, G.L. Payne, J.L. Friar, and B.F. Gibson,
Phys. Rev. C \textbf{33,} 1740 (1986).}

\bibitem{li96} {\small B.A. Li, Z.Z. Ren, C.M. Ko, and S.J. Yennello, Phys.
Rev. Lett. \textbf{76}, 4492 (1996); B.A. Li and A.T. Sustich, \textit{ibid.}
\textbf{82}, 5004 (1999).}

\bibitem{lgq9394} {\small G.Q. Li and R. Machleidt, Phys. Rev. C \textbf{48}%
, 1702 (1993); \textit{ibid.} \textbf{49} (1994) 566.}

\bibitem{pan93} {\small Q. Pan and P. Danielewicz, Phys. Rev. Lett. \textbf{%
70}, 2062 (1993).}

\bibitem{zhang93} {\small J. Zhang, S. Das Gupta, and C. Gale, Phys. Rev. C 
\textbf{50}, 1617 (1994).}

\bibitem{buu} {\small G.F. Bertsch and S. Das Gupta, Phys. Rep. \textbf{160}%
, 189 (1988).}

\bibitem{hei00} {\small H. Heiselberg and M. Hjorth-Jensen, Phys. Rep. 
\textbf{328}, 237 (2000).}

\bibitem{Posk71} {\small A.M. Poskanzer, G.W. Butler, and E.K. Hyde, Phys.
Rev. C \textbf{3}, 882 (1971).}

\bibitem{Lisa95} {\small M.A. Lisa \textit{et al.}, Phys. Rev. Lett. \textbf{%
75}, 2662 (1995).}

\bibitem{Poggi95} {\small G. Poggi \textit{et al.}, Nucl. Phys. \textbf{A586}%
, 755 (1995).}

\bibitem{Marie97} {\small N. Marie \textit{et al.}, Phys. Lett. \textbf{B391}%
, 15 (1997).}

\bibitem{Viola99} {\small V.E. Viola, K. Kwiatkowsky, and W.A. Friedman,
Phys. Rev. C \textbf{59}, 2660 (1999).}

\bibitem{Hagel00} {\small K. Hagel \textit{et al.}, Phys. Rev. C \textbf{62}%
, 034607 (2000).}

\bibitem{CorLong} {\small L.W. Chen, V. Greco, C.M. Ko, and B.A. Li, Phys.
Rev. C \textbf{68}, 014605 (2003); nucl-th/0305036.}

\bibitem{Naga81} {\small S. Nagamiya \textit{et al.}, Phys. Rev. C \textbf{24%
}, 971 (1981).}

\bibitem{Cibor00} {\small J. Cibor \textit{et al.}, Phys. Lett. B \textbf{473%
}, 29 (2000).}

\bibitem{Sobotka01} {\small L.G. Sobotka, R.J. Charity, and J.F. Dempsey, in %
\cite{ibook}, p.331.}

\bibitem{Vesel01} {\small M. Veselsky \textit{et al.}, Phys. Lett. B \textbf{%
497}, 1 (2001).}

\bibitem{Chomaz99} {\small Ph. Chomaz and F. Gulminelli, Phys. Lett. B 
\textbf{447}, 221 (1999).}

\bibitem{Gupta01} {\small S. Das Gupta and S.K. Samaddar, in \cite{ibook},
p.109.}

\bibitem{Boal90} {\small D.H. Boal, C.K. Gelbke and B.K. Jennings, Rev. Mod.
Phys. \textbf{62}, 553 (1990).}

\bibitem{bauer} {\small W. Bauer, C.K. Gelbke, and S. Pratt, Ann. Rev. Nucl.
Part. Sci. \textbf{42}, 77 (1992).}

\bibitem{ardo97} {\small D. Ardouin, Int. Jour. Phys. E \textbf{6}, 391
(1997).}

\bibitem{Wied99} {\small U.A. Wiedemann and U. Heinz, Phys. Rep. \textbf{319}%
, 145 (1999).}

\bibitem{gong90} {\small W. G. Gong et al., Phys. Rev. Lett. \textbf{65},
2114 (1990).}

\bibitem{gong91} {\small W.G. Gong, W. Bauer, C.K. Gelbke, and S. Pratt,
Phys. Rev. C \textbf{43}, 781 (1991).}

\bibitem{kunde93} {\small G. J. Kunde \textit{et al.}, Phys. Rev. Lett. 
\textbf{70}, 2545 (1993).}

\bibitem{handzy95} {\small D.O. Handzy \textit{et al.}, Phys. Rev. Lett. 
\textbf{75}, 2916 (1995).}

\bibitem{Verde02} {\small G. Verde, D.A. Brown, P. Danielewicz, C.K. Gelbke,
W.G. Lynch, and M.B. Tsang, Phys. Rev. C \textbf{65}, 054609 (2002).}

\bibitem{Verde03} {\small G. Verde, P. Danielewicz, D.A. Brown, W.G. Lynch,
C.K. Gelbke, and M.B. Tsang, Phys. Rev. C \textbf{67}, 034606 (2003) }

\bibitem{Ghetti00} {\small R. Ghetti \textit{et al.}, Nucl. Phys. \textbf{%
A674}, 277 (2000).}

\bibitem{Ghetti01} {\small R. Ghetti \textit{et al.}, Phys. Rev. Lett. 
\textbf{87}, 102701 (2001).}

\bibitem{Chitwood85} {\small C.B. Chitwood \textit{et al.}, Phys. Rev. Lett. 
\textbf{54}, 302 (1985).}

\bibitem{Poch85} {\small J. Pochodzalla\ \textit{et al.}, Phys. Rev. Lett. 
\textbf{55}, 177 (1985); J. Pochodzalla\ \textit{et al.}, Phys. Lett. 
\textbf{B174}, 36 (1986); J. Pochodzalla\ \textit{et al.}, Phys. Rev. C%
\textbf{35}, 1695 (1987).}

\bibitem{Kryger90} {\small R.A. Kryger \textit{et al.}, Phys. Rev. Lett. 
\textbf{65}, 2118 (1990); R.A. Kryger \textit{et al.}, Phys. Rev. C \textbf{%
46}, 1887 (1992).}

\bibitem{Zhu91} {\small F. Zhu \textit{et al.}, Phys. Rev. C \textbf{44},
R582 (1991).}

\bibitem{gong93} {\small W.G. Gong \textit{et al.}, Phys. Rev. C \textbf{47}%
, R429 (1993).}

\bibitem{Eraz94} {\small B. Erazmus, L. Martin, R. Lednicky, and N. Carjan,
Phys. Rev. C \textbf{49}, 349 (1994).}

\bibitem{Hamilton96} {\small T.M. Hamilton \textit{et al.}, Phys. Rev. C 
\textbf{53}, 2273 (1996).}

\bibitem{Kotte99} {\small R. Kotte \textit{et al.} (The FOPI Collaboration),
Eur. Phys. J. A \textbf{6}, 185 (1999).}

\bibitem{Gourio00} {\small D. Gourio \textit{et al.}, Eur. Phys. J. A 
\textbf{7}, 245 (2000).}

\bibitem{Ghetti03} {\small R. Ghetti et al., in XL Int. Winter Meeting on
Nucl. Phys., Bormio (Italy), 27 Jan-2 Feb, 2003}

\bibitem{pawel} {\small P. Danielewicz, Nucl. Phys. \textbf{A685}, 368c
(2001).}

\bibitem{Gelder95} {\small C.J. Gelderloos \textit{et al.}, Phys. Rev. Lett. 
\textbf{75}, 3082 (1995); C.J. Gelderloos \textit{et al.}, Phys. Rev. C 
\textbf{52}, R2834 (1995)}

\bibitem{koonin77} {\small S.E. Koonin, Phys. Lett. B \textbf{70}, 43 (1977).%
}

\bibitem{pratt1} {\small S. Pratt, Phys. Rev. Lett. \textbf{53}, 1219
(1984); Phys. Rev. D\textbf{33}, 72 (1986).}

\bibitem{pratt2} {\small S. Pratt and M.B. Tsang, Phys. Rev. C \textbf{36},
2390 (1987).}

\bibitem{hbt} {\small S. Pratt, Nucl. Phys. \textbf{A566}, 103c (1994).}

\bibitem{jacak} {\small B.V. Jacak \textit{et al.}, Phys. Rev. C \textbf{35}%
, 1751 (1987).}

\bibitem{borderie} {\small B. Borderie \textit{et al.}, Phys. Lett. \textbf{%
B388}, 224 (1996).}

\bibitem{greco99} {\small V. Greco, A. Guarnera, M. Colonna, and M. Di Toro,
Phys. Rev. C \textbf{59}, 810 (1999).}

\bibitem{das} {\small C. B. Das, S. Das Gupta, C. Gale, and B.A. Li, Phys.
Rev. C \textbf{67}, 034611 (2003).}

\bibitem{ono02} {\small A. Ono, Prog. Theor. Phys. Suppl. \textbf{146}, 378
(2002).}

\bibitem{bali03} {\small B. A. Li, C. B. Das, S. Das Gupta, and C. Gale,
submitted for publication.}
\end{thebibliography}
\end{document}